\documentclass[a4paper]{aa}
\usepackage{graphicx}

\def \sax {BeppoSAX}
\def \src {M33~X-8}
\def \degmark{^\circ}
\def \nh {N${\rm _H}$}

\def \hcm {\hbox {\ifmmode $ atom cm$^{-2}\else atom cm$^{-2}$\fi}}
\def \arcmin {\hbox{$^\prime$}}
\def \arcsec {\hbox{$^{\prime\prime}$}}

\def\approxgt{\mathrel{\hbox{\rlap{\lower.55ex \hbox {$\sim$}}
        \kern-.3em \raise.4ex \hbox{$>$}}}}
\def\approxlt{\mathrel{\hbox{\rlap{\lower.55ex \hbox {$\sim$}}
        \kern-.3em \raise.4ex \hbox{$<$}}}}
\newcommand{\mc}{\multicolumn}

\begin{document}

\thesaurus{(11.09.1; 11.14.1; 13.25.2)}  

\title{BeppoSAX spectroscopy of the luminous X-ray sources in M33}

\author{A.N. Parmar\inst{1}
       \and L. Sidoli\inst{1} 
       \and T. Oosterbroek\inst{1}
       \and P.A. Charles\inst{2}
       \and G.~Dubus\inst{3}
       \and M.~Guainazzi\inst{4}
       \and P.~Hakala\inst{5}
       \and W.~Pietsch\inst{6}
       \and G.~Trinchieri\inst{7}
}
\offprints{A.N. Parmar (aparmar@astro.estec. esa.nl)}

\institute{
        Astrophysics Division, Space Science Department of ESA, ESTEC,
        Postbus 299, 2200 AG Noordwijk, The Netherlands
\and
        Dept of Physics \& Astronomy, University of Southampton,
        Southampton, Hants, SO17 1BJ, United Kingdom
\and
        Astronomical Institute ``Anton Pannekoek'', Kruislaan 403,
        1098 SJ Amsterdam, The Netherlands
\and   
        XMM-Newton SOC, ESA Villafranca, Apartado 50727, E-28080, Madrid
        Spain
\and
        Observatory \& Astrophysics Lab., University of Helsinki,
        Helsinki, Finland
\and
        Max-Planck-Institut f\"ur Extraterrestrische Physik, D-85740 
        Garching bei M\"unchen, Germany
\and
        Osservatorio Astronomico di Brera, Via Bianchi 46, Merate, I-22055
        Italy
}
\date{Received ; accepted }

\maketitle

\markboth{BeppoSAX Observations of M33}{BeppoSAX Observations of M33}

\begin{abstract}
The nearby galaxy M33 was observed by the imaging X-ray instruments on-board
BeppoSAX. Two observations at different phases of the 105.9~day 
intensity cycle of the luminous central source X-8 failed
to reveal the expected modulation, suggesting that it is
probably transitory. Similar behavior has been observed from
several X-ray binary sources. This strengthens somewhat the 
idea that \src\ is a black hole accreting from a binary
companion.
The 0.2--10~keV spectrum of \src\ can best be 
modeled by an
absorbed power-law with a photon index, $\alpha$, of 
$1.89 \, \pm \, ^{0.40} _{0.79}$
and a disk-blackbody with a temperature, kT, of $1.10 \pm 0.05$~keV and
a projected inner-disk radius of $55.4 \, \pm \, ^{6.0} _{7.7}$~km. 
This spectral shape is in good agreement with earlier ASCA results.
The 2--10~keV spectra of M33~X-4, X-5, X-7, X-9 and X-10 are all consistent 
with power-law or bremsstrahlung models, whilst that of X-6 appears to be
significantly more complex and may be reasonably well modeled with a
disk-blackbody with kT = $1.7 \pm 0.2$~keV and
a projected inner-disk radius of $7 \pm 2$~km. The spectrum
of M33~X-9 is rather hard with $\alpha = 1.3$. 
Compared to earlier $Einstein$ and ROSAT observations, 
M33~X-7 and X-9 varied in intensity, M33~X-4, X-6, and X-10
may have varied and M33~X-5 remained constant.


\end{abstract}

\section{Introduction}
\label{sect:intro}

\begin{table*}
\caption{BeppoSAX M33 observing log. Phase, $\Phi$, is determined using the
105.9~day ephemeris of Dubus et al. (\cite{d:97}) 
where the \src\ maximum is expected at phase 0.0 and minimum at $\sim$0.7}
\begin{tabular}{llrrrccc}
\hline\noalign{\smallskip}
Num. & \mc{2}{c}{Observation} & \mc{2}{c}{Exposure}& \mc{2}{c}{\src\ 
count rate}
&$\Phi$\\
       &  \mc{1}{c}{Start} & \mc{1}{c}{End} & LECS & MECS & LECS  & MECS \\
       & (year~mn~dy~hr:mn) & (mn~dy~hr:mn) & (ks) &(ks)
&(0.1--2 keV; s$^{-1}$)& (1.8--10~keV; s$^{-1}$) \\
\noalign {\smallskip}
\hline\noalign {\smallskip}
1 & 1998~Aug~24 13:05 & Aug~25 22:26 & 32.3 & 59.9 & $0.068 \pm 0.002$ &
$0.157 \pm 0.002$ & 0.72\\
2 & 1999~Jan~06 20:16 & Jan~07 18:38 &  0.9 &  1.1 & $0.045 \pm 0.008$  &
$0.131 \pm 0.011 $  & 0.98 \\
3 & 2000~Jul~09 17:56 & Jul~13 04:33 &  7.3 & 37.4 & $0.065 \pm 0.004$  &
$0.141 \pm 0.002 $  & 0.20 \\
\noalign {\smallskip}
\hline
\end{tabular}
\label{tab:observing_log}
\end{table*}

M33 (NGC 598) is one of the nearest galaxies 
(795~kpc, van den Bergh \cite{vdb:91}) and is 
classified as a late-type Sc spiral. 
As well as containing several point X-ray sources, M33 is unique
amongst the Local Group
in having a very bright ($\sim$10$^{39}$ erg s$^{-1}$)
X-ray source (X-8) located at a position consistent (to within
the positional error of the {\it Einstein} High Resolution Imager
(HRI) of 3\arcsec) with the optical
nucleus of the galaxy (see Long et al. \cite{l:81}; Markert \&
Rallis \cite{m:83};
Gottwald et al. \cite{g:87}; Trinchieri et al. \cite{t:88}; 
Dubus et al. \cite{d:97}). 
This source dominates
the X-ray flux from the galaxy, being responsible for $\sim$70\% of the
0.15--4.5~keV emission. 
Another X-ray source in M33 (X-7) is an eclipsing X-ray binary 
pulsar (Peres et al. \cite{p:89}; 
Schulman et al. \cite{s:93}, \cite{s:94}; Larson \& Schulman \cite{ls:97})
with a 3.45~day orbital period 
and a 0.31~s pulse period (Dubus et al. \cite{d:99}).
The X-ray sources in M33 have been studied by {\it Einstein}, EXOSAT,
ROSAT, and ASCA, as well as by earlier non-imaging instruments.
With an exposure of 50~ks ROSAT Position Sensitive Proportional
Counter (PSPC) observations of M33 
have detected 35 sources in the central 30$'$
which are brighter than 10$^{36}$ erg s$^{-1}$ (Long et al. \cite{l:96};
see also Haberl \& Pietsch \cite{hp:00}).
From an examination of archival ROSAT data spanning over 6~years
(a total of 387~ks of HRI time and 62~ks of PSPC time),
Dubus et al. (\cite{p:97}) find that the X-ray intensity of \src\ 
is almost certainly modulated with a period of $105.9 \pm 0.1$~days.

The nature of \src\ is unclear. X-ray transients
such as Aql~X-1 and 4U-1630-47 show outbursts on
timescales of hundreds of days, but are not periodic (e.g., 
Kuulkers et al. \cite{k:97}) while a number of persistent
X-ray sources exhibit long term periodicities in their 
X-ray intensities. Extended observations using the All-Sky
Monitor (ASM) on the Rossi X-ray Timing Explorer (R-XTE) have 
revealed changes in some long term periodicities (e.g., 
Paul et al. \cite{p:00}); Wojdowski et al.
\cite{w:98}; Kong et al. \cite{k:98}).
The detection of a 106~day periodicity over $\sim$20 cycles
implies a regularity and duty cycle incompatible
with typical soft X-ray transients. 
\src\ may instead be a moderate mass black hole.
Until recently, the M33 nuclear source was frequently 
considered (as are many of
the bright sources in normal galaxies that exceed the Eddington limit of
$\sim$10$^{38}$ erg s$^{-1}$ for a 1M$_\odot$ compact object) to be a ``mini-''
(or perhaps ``micro-'') active galactic nucleus (AGN).
However, the X-ray spectrum of X-8 appears to be much steeper
than a classical AGN power-law and can be well fit by
a multi-color disk blackbody model plus a harder 
power-law (Takano et al. \cite{t:94}).
This spectral shape is characteristic of the galactic black hole candidates
which is often ultra-soft at 
low-energies and ultra-hard at high-energies. The ultra-soft
component may originate from an optically thick accretion disk,
while the ultra-hard component may result from comptonization
of soft photons by hot electrons (Sunyaev \& Tr\"umper \cite{s:79}).
The accretion disk model is parameterized by only two quantities,
the projected inner disk radius, ${\rm r_{in}({\cos}i)^{0.5}}$, and 
the temperature at this inner radius, ${\rm kT_{in}}$
(Mitsuda et al. \cite{m:84}; Makishima et al. \cite{m:86}).
Using ASCA, Takano et al. (\cite{t:94}) 
obtain a value for ${\rm r_{in}({\cos}i)^{0.5}}$ of $\sim$50~km, typical of 
black hole candidates in their high state. At the distance of M33,
this corresponds to a black hole mass of $\sim$10~M$_\odot$. 

In this {\it paper} we report on BeppoSAX observations of M33 primarily
designed to investigate any spectral changes associated with the 105.9~day
cycle of \src\ (Sect.~\ref{sect:m33x-8}) as well as the spectral
and temporal properties of the
luminous X-ray sources in M33 (Sect.~\ref{sect:pointsources}).

\section{Observations}
\label{sect:obs}

Data from the coaligned Low-Energy Concentrator Spectrometer (LECS;
0.1--10~keV; Parmar et al. \cite{p:97}), Medium-Energy Concentrator
Spectrometer (MECS; 1.8--10~keV; Boella et al. \cite{b:97}), and
the Phoswich Detection System (PDS; 15--300~keV; Frontera et al. \cite{f:97})
on-board BeppoSAX are presented. 
The MECS consists of two identical grazing incidence
telescopes with imaging gas scintillation proportional counters in
their focal planes. The LECS uses an identical concentrator system as
the MECS, but utilizes an ultra-thin entrance window and
a driftless configuration to extend the low-energy response to
0.1~keV. The fields of view (FOV) of the LECS and MECS are circular
with diameters of 37\arcmin\ and 56\arcmin, respectively.  
The MECS has a 2\arcmin\ thick circular window support structure
which is centered 10\arcmin\ from the FOV center. In addition,
four 2\arcmin\ thick radial spokes extend outwards from the circular
support. The non-imaging
PDS consists of four independent units arranged in pairs each having a
separate collimator. Each collimator was alternatively
rocked on- and 210\arcmin\ off-source every 96~s during the observations.

M33 was observed 3 times by \sax\
(see Table~\ref{tab:observing_log}). The observing
times were chosen in order to study \src\ near
the maximum and minimum of the 105.9~day X-ray cycle. The maxima
occur at MJD = 48750.3 + n 105.9, and the minima approximately 0.7
in phase later (Dubus et al. \cite{d:97}). The 1998 August observation
was close to the 105.9~day minimum ($\Phi = 0.72$), 
and the 1999 January observation close to the maximum ($\Phi = 0.98$). 
However, this observation was curtailed due
to technical problems and repeated in 2000 July at 
$\Phi = 0.20$, still close to the expected maximum of the 105.9~day
cycle. Good data were selected from intervals when the elevation angle
above the Earth's limb was $>$$5^{\circ}$ and when the instrument
configurations were nominal, using the SAXDAS 2.0.0 data analysis package.

\begin{figure*}
  \centerline{\hbox{\hspace{-1.5cm}
         \includegraphics[height=13.5cm,angle=0.0]{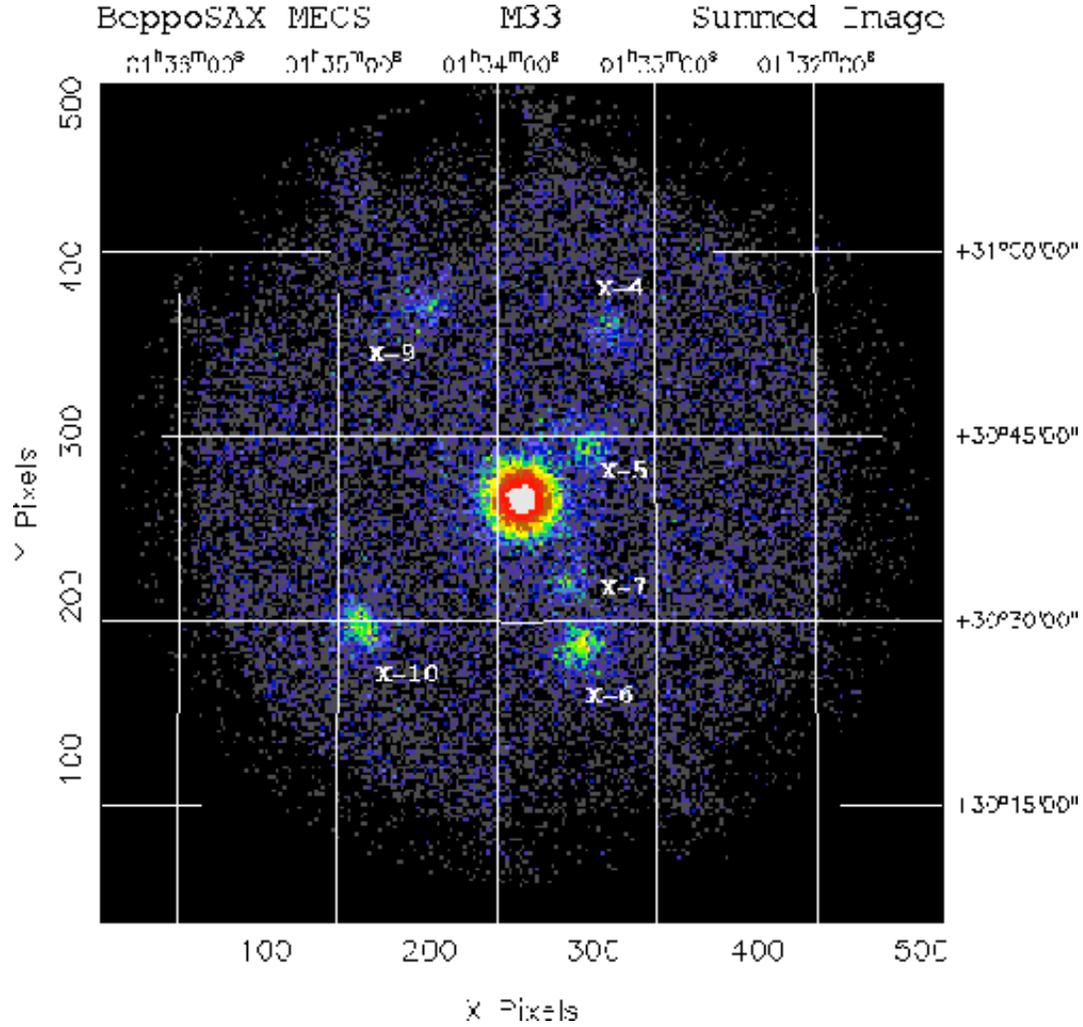}}}
%  \vspace{-1.0cm}
  \caption[]{MECS 1.8--10.0 keV image of M33 obtained by summing data
   from all 3 observations. \src\ is the central bright source. 
  The data have been rebinned by a factor 2 into 16\arcsec\ pixels.
  The location of the sources discussed in Sect.~\ref{sect:pointsources}
  are indicated. The ``cut-outs'' at the top and bottom of the image are
  due to the removal of instrument calibration source events}
  \label{fig:images}
\end{figure*}

Fig.~\ref{fig:images} shows the MECS 1.8--10~keV image obtained by summing
data from all 3 observations. The total exposure is 98.3~ks. 
For clarity, the data have been rebinned by a factor 2 into 
16\arcsec\ pixels. No corrections for vignetting or obscuration
by the MECS window support structure (see above) have been applied. 
As well as the central \src\ source which dominates the overall X-ray
flux, M33 X-4, X-5, X-6, X-7, 
X-9 and X-10 are clearly detected. The luminous
source M33~X-3 is strongly obscured by the MECS
strongback and was not detected. In addition, the source identified
here as M33~X-9 is resolved into 3 sources in ROSAT and 
{\it Einstein}
observations (Long et al. \cite{l:96}; Peres et al. \cite{p:89}).

\begin{table*}
\caption[]{MECS positions (J2000) of the faint point sources  
($\sim$1$'$ uncertainty radius),
compared with the ROSAT PSPC positions reported in Long et al. (\cite{l:96}). 
Source X-9 is unresolved in the MECS observations, but resolved into three 
source (a,b,c) with ROSAT and $Einstein$. For X-9 (Long et al. \cite{l:81}) 
and X-9(c) (Peres et al. \cite{p:89}) the $Einstein$ coordinates are
given. B74=Boulesteix et al. \cite{b:74} catalog of M33 H~{\sc ii} regions; 
L81=Long et al. \cite{l:81}; L96=Long et al. \cite{l:96}; 
CS82a=Christian \& Schommer \cite{cs:82a}; 
CS82b=Christian \& Schommer \cite{cs:82b}}
\begin{tabular}{lccccl}
\hline\noalign {\smallskip}
Source 		        &   R.A. (MECS)       & Dec. (MECS)    &  R.A. (PSPC)       & Dec. (PSPC)  &  Notes  \\
 			& \mc{1}{c}{(h~m~s)}  & \mc{1}{c}{($\degmark$ $\arcmin$ $\arcsec$)}   
                        & \mc{1}{c}{(h~m~s)}  & \mc{1}{c}{($\degmark$ $\arcmin$ $\arcsec$)}    &        \\
\noalign {\smallskip}
\hline\noalign {\smallskip}
X-4    & 01 33 17.5  & $+$30 54 02.0  & 01 33 14.8 & $+$30 53 28.5 & SNR 136+396 (L96); H~{\sc ii} IC133 (L81) \\
X-5    & 01 33 25.9  & $+$30 44 33.8  & 01 33 24.3 & $+$30 44 04.7 & \dots \\
X-6    & 01 33 27.7  & $+$30 28 12.5  & 01 33 27.5 & $+$30 27 22.7 & Globular cluster? (L81; CS82a)  \\
X-7    & 01 33 31.8  & $+$30 33 36.4  & 01 33 33.8 & $+$30 32 12.8 & X-ray eclipsing binary; H~{\sc ii} 
B208 (B74; L81)  \\
X-9    & 01 34 26.0  & $+$30 55 59.7  &  01 34 36.3 & $+$30 55 00.4 & $Einstein$ position (L81) \\
X-9(a) &  \dots      &   \dots      & 01 34 38.7 & $+$30 55 10.3 &  \dots\\
X-9(b) &  \dots      &   \dots      & 01 34 25.5 & $+$30 55 24.5 & Globular cluster? (n.3/4 in CS82b)  \\
X-9(c) &  \dots      &   \dots      & 01 34 31.5 & $+$30 56 20.5 & Elliptical galaxy at z=0.03? (CS82a) \\
X-10   & 01 34 50.8  & $+$30 29 17.3  & 01 34 51.7 & $+$30 29 05.9  & \dots  \\
\noalign {\smallskip}
\hline
\end{tabular}
\label{tab:positions}
\end{table*}

\begin{table}
\caption[]{Point sources detected in the MECS. C is 
the average MECS 1.8--10~keV background subtracted count rate. 
R is the extraction radius used for the MECS spectral analysis and
$\theta$ the source off-axis angle. Note that the MECS strongback
obscures sources located between $\theta = 9$--11\arcmin. V is the
approximate vignetting correction to correct C to the on-axis case}
\begin{tabular}{lllrr}
\hline\noalign {\smallskip}
Source   &  \mc{1}{c}{C}        & R  &  $\theta$  & V \\
         & (10$^{-3}$~s$^{-1})$ &  ($\arcmin$) &  ($\arcmin$) \\
\noalign {\smallskip}
\hline\noalign {\smallskip}
X-4      & $2.61 \pm 0.26$       & 4 & 16  & 2.4 \\
X-5      & $3.40 \pm 0.26$       & 2 &  6  & 1.3 \\
X-6      & $6.15 \pm 0.29$       & 2 & 12  & 1.9 \\
X-7      & $2.07 \pm 0.19$       & 2 &  7  & 1.3 \\
X-8      & $151.8 \pm 1.3$       & 4 &  0  & 1.0 \\
X-9      & $2.85 \pm 0.27$       & 4 & 18  & 2.7 \\
X-10     & $8.95 \pm 0.40$       & 4 & 17  & 2.5 \\
\noalign {\smallskip}
\hline
\end{tabular}
\label{tab:pointsources}
\end{table}

\section{\src\ analysis}
\label{sect:m33x-8}

LECS and MECS data were extracted centered on the position of \src\ 
using radii of 6\arcmin\ and 4\arcmin, respectively. This LECS
extraction radius is smaller than normally used in order to minimize the
contribution from nearby sources. The cosmic and internal instrument
backgrounds are $\sim$5\% of the source flux, and so background
subtraction is not critical and was performed using source free 
regions of sky. 
Table~\ref{tab:observing_log} gives the 0.1--2.0~keV (LECS) and 2.0--10~keV
(MECS) count rates for each observation. 

The \src\ spectra were investigated by simultaneously
fitting data from the LECS and MECS instruments. 
All spectra were rebinned to oversample the full
width half maximum of the energy resolution by
a factor 3 and to have additionally a minimum of 20 counts 
per bin to allow use of the $\chi^2$ statistic. Data
were selected in the energy ranges
0.2--8.0~keV (LECS) and 1.8--10~keV (MECS) 
where the instrument responses are well determined and sufficient
counts obtained. 
The photoelectric absorption
cross sections of Morrison \& McCammon (\cite{mc:83}) 
are used throughout.
Factors were included in the spectral fitting to allow for normalization 
uncertainties between the two instruments. 

\subsection{1998 August spectrum}

\begin{table*}
\caption[]{\src\ fit results.
\nh\ is the absorption in units of $\rm {10^{22}}$ atom $\rm {cm^{-2}}$.
${\rm r_{in}({\cos}i)^{0.5}}$ is in units of km for a distance
of 795~kpc. $\alpha$ is the power-law photon index.
90\% confidence limits are given when the $\chi ^2$ per dof is $<$2}
\begin{flushleft}
\begin{tabular}{lcccccr}
\hline\noalign{\smallskip}
Model & \hfil N$_{\rm {H}}$ \hfil & kT/kT${\rm _{in}}$ (keV) &$\alpha$
& $\rm{E_{co}}$ (keV) & ${\rm r_{in}({\cos}i)^{0.5}}$ & $\chi^2$/dof \\
\noalign{\smallskip\hrule\smallskip}
1998 August observation\\
\quad Power-law & 0.58 & \dots & 2.7 & \dots & \dots
& 380.8/101  \\
\quad Bremsstrahlung & $0.24 \pm 0.03$ & $2.93 \pm 0.10 $ & \dots & \dots &
\dots & 149.8/101\\
\quad Cutoff power-law & $0.12 \pm 0.04 $ & \dots &$ 0.63 \pm
0.29$ & $1.89 \pm 0.17$ & \dots & 118.1/100\\
\quad Power-law + blackbody & $0.23 \pm 0.05$ & $0.75 \pm
0.03$ & $ 2.44 \pm 0.15$ & \dots & \dots & 106.0/99 \\
\quad Power-law + disk-blackbody & $0.15 \pm 0.06$ & $1.12 \pm
0.06$ & $ 2.15 \, \pm \,_{0.70} ^{0.40}$ & \dots & $55.2 \, 
\pm \, ^{7.6} _{5.7}$
& 108.9/99 \\
\noalign{\smallskip\hrule\smallskip}
2000 July observation\\
\quad Power-law & $0.55 \pm 0.10$ & \dots & $2.60 \pm 0.07$ & \dots & \dots
& 105.3/71  \\
\quad Bremsstrahlung & $0.21 \pm 0.06$ & $3.40 \pm 0.20 $ & \dots & \dots &
\dots & 75.4/71\\
\quad Cutoff power-law & $0.25 \pm 0.10 $ & \dots &$ 1.49 \, \pm \, 
^{0.33} _{0.08}$ & $3.79 \, \pm \, ^{1.68} _{0.93}$ & \dots & 75.2/70\\
\quad Power-law + blackbody & $0.29 \pm 0.15$ & $0.72 \, \pm \, ^{0.14}
_{0.07}$ & $ 2.33 \pm 0.20$ & \dots & \dots & 76.9/69 \\
\quad Power-law + disk-blackbody & $0.20 \pm 0.20$ & $1.05 \, \pm \, ^{0.21}
_{0.10}$ & $ 2.06 \, \pm \, _{0.73} ^{0.51}$ & \dots 
& $50.0 \, \pm \, ^{6.0} _{7.7}$
& 74.0/69 \\
\noalign{\smallskip\hrule\smallskip}
All observations\\
\quad Power-law & 0.55& \dots & 2.7 & \dots & \dots
& 396.6/105 \\
\quad Bremsstrahlung & $0.22 \pm 0.03$ & $3.10 \pm 0.09 $ & \dots & \dots &
\dots & 141.9/105\\
\quad Cutoff power-law & $0.135 \pm 0.05 $ & \dots &$ 0.85 \pm
0.18$ & $2.22 \, \pm \, ^{0.25} _{0.20}$ & \dots & 125.4/104\\
\quad Power-law + blackbody & $0.25 \pm 0.05$ & $0.75 \pm 0.03
$ & $ 2.40 \pm 0.12$ & \dots & \dots & 122.2/103 \\
\quad Power-law + disk-blackbody & $0.12 \, \pm \, ^{0.10} _{0.06}$ 
& $1.10 \pm 0.05$ & $ 1.89 \, \pm \, _{0.79} ^{0.40}$ & \dots & 
$55.4 \, \pm \, ^{6.0} _{7.7}$ & 115.7/103 \\
\noalign{\smallskip\hrule\smallskip}
\end{tabular}
\end{flushleft}
\label{tab:spec_paras}
\end{table*}

Initially, simple models were tried, including absorbed power-law,
thermal bremsstrahlung and cutoff power-law 
(${\rm E^{-\alpha}\exp-(E_{co}/kT)}$) models. 
The power-law and bremsstrahlung models 
give unacceptable fits with $\chi ^2$ of 380.8 and 149.8 for 101 degrees 
of freedom (dof), respectively.
The cutoff power-law model fit is acceptable with a $\chi ^2$ of
118.1 for 100 dof. Next, more complex models consisting of a power-law
and a blackbody and a power-law and a disk-blackbody were fit. 
These both give acceptable fits with
$\chi ^2$s of 106.0 and 108.9 for 99 dof, respectively. The
best-fit inner disk radius is $55.2 \, \pm \, ^{7.6} _{5.7}$~km for a
distance of 795~kpc and a disk inclination of 0$\degmark$.
The fit results are summarized in Table~\ref{tab:spec_paras}.
The absorption corrected 1.0--10~keV source luminosity is 
$1.3 \times 10^{39}$~erg~s$^{-1}$ for a distance of 795~kpc.

\subsection{2000 July spectrum}

The spectra were created in the same way as for the 1998 August observation and
the same spectral models applied. The results are given in 
Table~\ref{tab:spec_paras}. The exposure times for this observation are
shorter than for the 1988 August observation, and the spectrum can be
successfully modeled using either bremsstrahlung, cutoff power-law, 
power-law
and blackbody or power-law and disk-blackbody models. The derived spectral
parameters are consistent with those from the earlier observation with an
inner disk radius of $50.0 \, \pm \, ^{13.2} _{22.6}$~km for the 
same assumptions
as above. The absorption corrected
1.0--10~keV source luminosity is $1.2 \times 10^{39}$~erg~s$^{-1}$
for a distance of 795~kpc.

\subsection{Combined spectrum}

Since the spectral fit parameters derived from the fits to the 1998 
August and 2000 July data are consistent, these data were combined to 
produce a ``summed'' spectrum. Data from the short
(MECS exposure of 1.1~ks) observation in 1999 January were also included.
The same spectral models as above were fit and the results summarized in
Table~\ref{tab:spec_paras}. As found by Takano et al. (\cite{t:94}), the
best-fit model is the power-law and disk-blackbody combination with a
${\rm kT_{in}}$ of $1.10 \pm 0.05$~keV and an inner radius of 
$55.4 \, \pm \, ^{6.0}
_{7.7}$~km for a distance of 795~kpc and a disk inclination angle of 
0$\degmark$. The mean, absorption corrected, 1.0--10~keV luminosity of 
\src\ is $1.25 \times 10^{39}$~erg~s$^{-1}$ for a distance of 795~kpc.
The 90\% confidence upper limit to a narrow emission line at 6.5~keV is
100~eV.

The galactic absorption in the direction of \src\ is 
$5.6 \times 10^{20}$~atom~cm$^{-2}$ (Dickey \& Lockman \cite{d:90}).
Using the power-law and disk-blackbody model, the best fit absorption
of $(1.2 \, \pm \, ^{1.0} _{0.6}) \times 10^{21}$~atom~cm$^{-2}$ suggests
the presence of a small amount of additional absorption. The 3$\sigma$
upper-limit to any PDS counts in the 15--40~keV band is 0.06~s$^{-1}$,
consistent with the extrapolation of the best-fit power-law and 
disk-blackbody spectrum to this energy range which gives 
0.02~count~s$^{-1}$.
For a power-law spectrum with $\alpha = 2.1$ the upper-limit
corresponds to a flux of $<$0.4~mCrab. 

\begin{figure*}
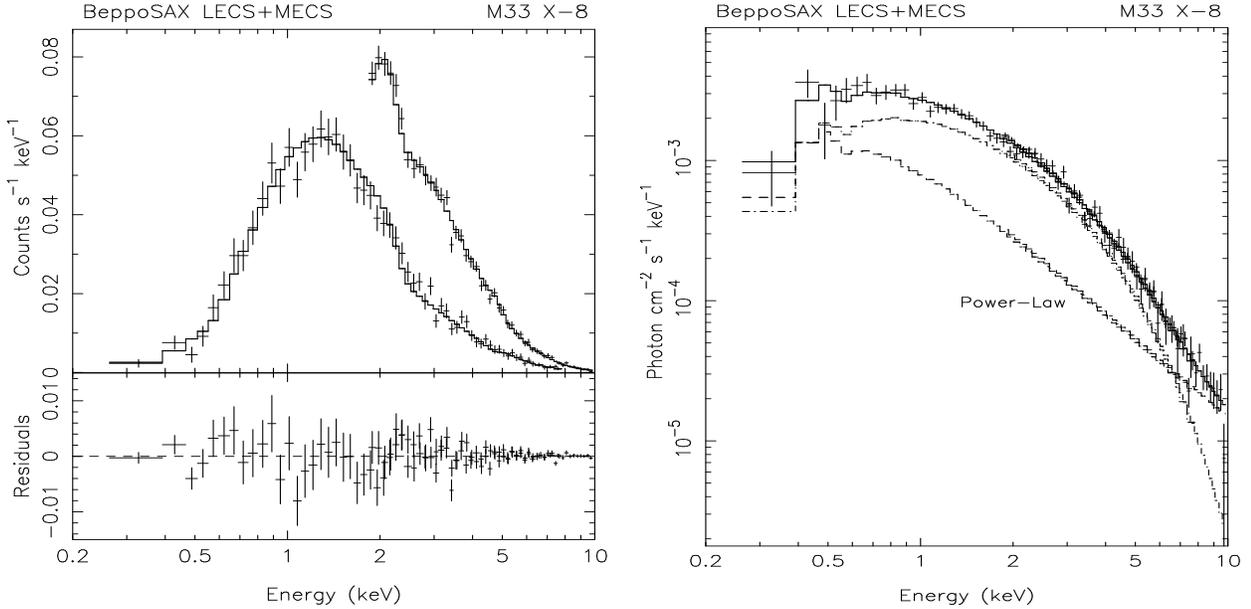

  \hbox{\hspace{0.4cm}
  \includegraphics[height=8.0cm,width=8.0cm,angle=-90]{h2503f2a.ps}
  \hspace{0.2cm}
  \includegraphics[height=8.0cm,width=8.0cm,angle=-90]{h2503f2b.ps}}
  \caption[]{The \src\ LECS and MECS count spectrum and residuals when the 
             power-law and disk-blackbody model is fit to the summed 
             spectrum (left
             panel). The residuals are given in count~s$^{-1}$~keV$^{-1}$.
             The right panel shows the photon spectrum with the contribution
             of the power-law indicated}
  \label{fig:spectrum}
\end{figure*}

\subsection{Time Variability}
\label{subsect:timing}

\begin{figure}
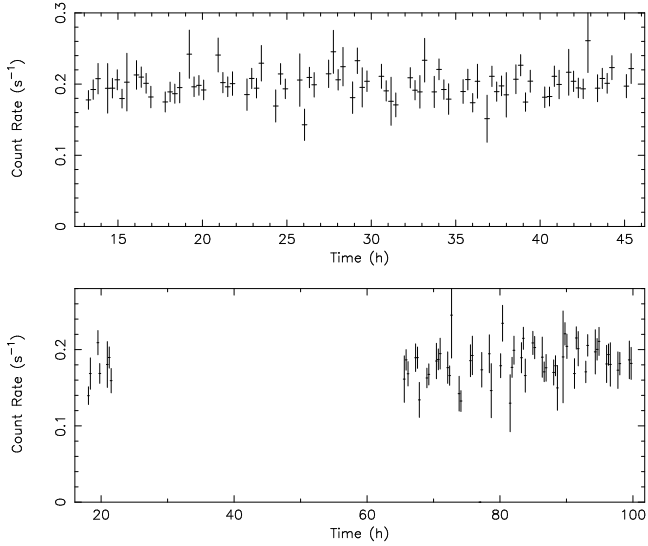

\hbox{\hspace{0.0cm}\includegraphics[height=8.5cm,angle=-90]{h2503f3a.ps}}
\vspace{0.25cm}
\hbox{\hspace{0.0cm}\includegraphics[height=8.5cm,angle=-90]{h2503f3b.ps}}
  \caption[]{MECS 2--10~keV lightcurves for \src\ with a binning
of 1024~s for the 1998 August (upper panel) and 2000 July (lower panel)
observations. No background has been subtracted}
  \label{fig:x8lc}
\end{figure}

Fig.~\ref{fig:x8lc} shows 2--10~keV MECS lightcurves of \src\ during
1998 August and 2000 July observations with a binning time of 1024~s.
The LECS and MECS data from these observations
were searched for the presence of periodic signals. None were found.
During the 1998 August observation the 3$\sigma$ limit to any 2--10~keV
(MECS) root mean square variability is $<$7.6\% when evaluated on a 
timescale
of 3000~s. In the energy band 0.1--2.0~keV (LECS) the upper
limit evaluated in the same way is $<$5.3\%. For the 2000 July observation 
the corresponding upper limits are $<$25\% and $<$29\%.

\begin{figure}
\includegraphics[height=7.5cm,angle=-90]{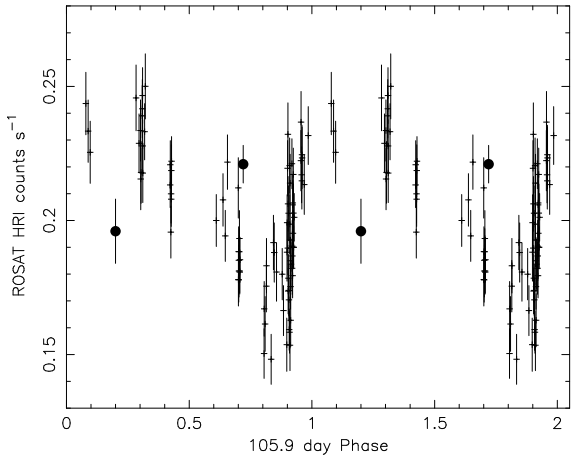}
  \caption[]{The ROSAT HRI count rates of Dubus et al. (\cite{d:97})
  folded over the 105.9 day period. The count rates corresponding to 
  the two BeppoSAX measurements are indicated with filled circles}
  \label{fig:folded_lc}
\end{figure}

The mean ROSAT HRI count rate is 0.20~s$^{-1}$ and the
modulation appears sinusoidal-like with an amplitude of $\sim$20\%.
The folded 0.1--2.5~keV ROSAT lightcurve (see Fig.~\ref{fig:folded_lc} and
Fig.~3 of Dubus et al. \cite{d:97}) has a mean count rate of 
0.20~s$^{-1}$ with a sinusoidal-like
modulation superposed with an amplitude of $\sim$20\%. This
implies that the {\it count rates} of the 1999 January and 2000 July 
observations should be $\sim$40\% higher than that
of 1998 August, should the 105.9~day modulation be present.
However, the MECS 2.0--10.0~keV count rate at $\Phi = 0.72$ 
(close to the expected {\it minimum} of the cycle) is higher than at the
expected {\it maximum}. In order to compare our BeppoSAX
results for the two long observations with those of Dubus et al. 
(\cite{d:97}), we folded the best-fit BeppoSAX power-law and
disk-blackbody models (see Table~\ref{tab:spec_paras})
through the HRI on-axis response matrix obtained from the HEASARC. 
This gives predicted HRI count rates of $0.221 \pm 0.007$~s$^{-1}$
and $0.195 \pm 0.012$~s$^{-1}$ for the 1998 August and 2000 July
observations, respectively. These points are shown together with the
HRI results of Dubus et al. (\cite{d:97}) on Fig.~\ref{fig:folded_lc}.
The derived count rates do not support the continued presence of the 
105.9~day modulation, although we cannot entirely exclude the 
possibility that the modulation is being masked by a higher than
usual level of intrinsic variability.

\section{Point sources}
\label{sect:pointsources}

\begin{table*}
\caption[]{Spectral fit results to the M33 point sources.
The absorption, ${\rm N_H}$, is in units of $10^{22}$~atom~cm$^{-2}$.
Uncertainties are quoted at 90\% confidence. 
The 2--10~keV absorption corrected source luminosity, L, assumes a 
distance of 795 kpc}
\label{tab:faint}
\begin{tabular}{llcccrr}
\hline\noalign{\smallskip}
Source          & Model & \mc{1}{c}{${\rm N_H}$}    
& \mc{1}{c}{kT} & \mc{1}{c}{$\alpha$}  & \mc{1}{c}{L}  &  $\chi^2$/dof \\
                &       &                &             
 \mc{1}{c}{(keV)}   &   &    (erg s$^{-1}$) &\\
\noalign{\smallskip\hrule\smallskip}
X-4  & Power-law      & $2.0 \, \pm \, ^{6.5} _{1.9}$  & \dots & $3.4 \, \pm \, ^{2.8} _{1.0}$ 
&  $1.5 \times 10^{38}$ & 13.7/12 \\
     & Power-law      & 0.056 fixed  & \dots & $2.7 \, \pm \, ^{0.6} _{0.4}$ 
&  $7.9 \times 10^{37}$ & 15.0/13 \\
     & Bremsstrahlung & $<$4.9 &  $1.8 \, \pm \, ^{1.8} _{0.9}$ &\dots 
& $8.6\times 10^{37}$& 13.2/12 \\
     & Bremsstrahlung & 0.056 fixed  &  $2.3 \, \pm \, ^{1.4} _{0.6}$ &\dots 
& $7.0\times 10^{37}$& 13.7/13 \\
\noalign{\smallskip\hrule\smallskip}
X-5  & Power-law      & $<$5.6  & \dots & $3.2 \, \pm \, ^{1.2}_{0.9}$ 
&  $3.4 \times 10^{37}$ & 10.7/14  \\
     & Power-law      & 0.056 fixed  & \dots & $2.6 \, \pm \, ^{0.4} _{0.4}$ 
&  $2.7 \times 10^{37}$ & 11.6/15 \\
     & Bremsstrahlung & $<$2.9 &  $2.9 \, \pm \, ^{1.5} _{1.2}$ &\dots 
& $2.6\times 10^{37}$& 9.7/14 \\
     & Bremsstrahlung & 0.056 fixed  &  $2.9 \, \pm \, ^{1.7} _{0.7}$ &\dots 
& $2.5\times 10^{37}$& 9.7/13 \\
\noalign{\smallskip\hrule\smallskip}
X-6  & Power-law      & $0.45 \, \pm \, ^{0.70} _{0.32}$ & \dots 
& $2.1 \, \pm \, ^{0.2} _{0.3}$ &  $7.6\times 10^{37}$  & 51.7/28   \\
     & Power-law      & 0.056 fixed & \dots 
& $1.8 \, \pm \, ^{0.1} _{0.1}$ &  $7.5\times 10^{37}$  & 56.9/29   \\
     & Bremsstrahlung & $<$0.69 &  $5.6 \, \pm \, ^{2.5} _{1.3}$ &\dots 
& $7.1\times 10^{37}$& 44.2/28 \\
     & Bremsstrahlung & 0.056 fixed &  $6.4 \, \pm \, ^{2.2} _{1.3}$ &\dots 
& $7.1\times 10^{37}$& 45.0/29 \\
     & Blackbody      & $<$0.19 & $1.0 \pm 0.10$ &  \dots &  
$7.1\times 10^{37}$  & 41.4/28   \\
     & Disk-blackbody & $<$0.29 & $1.7 \pm 0.2$ & \dots & 
$6.8 \times 10^{37}$ & 37.2/28 \\ 
\noalign{\smallskip\hrule\smallskip}
X-7  & Power-law & $<$9.8   &\dots & $2.9 \, \pm \, ^{1.7}_{1.3}$  
&  $2.8 \times 10^{37}$   & 6.2/9 \\
     & Power-law      & 0.056 fixed  & \dots & $1.7 \, \pm \, ^{0.6} _{0.6}$ 
&  $1.6 \times 10^{37}$ &7.3/10 \\
     & Bremsstrahlung & $<$11 &  $3.7 \, \pm \, ^{97} _{2.4}$ &\dots 
& $1.8\times 10^{37}$& 6.1/9 \\
     & Bremsstrahlung & 0.056 fixed  &  $>$3.7 &\dots 
& $1.4\times 10^{37}$& 6.9/10 \\
\noalign{\smallskip\hrule\smallskip}
X-9  & Power-law & $<$3.9   &\dots & $1.3 \, \pm \, ^{2.3}_{0.8}$  
&  $6.1 \times 10^{37}$  & 7.2/8 \\
     & Power-law      & 0.056 fixed  & \dots & $1.1 \, \pm \, ^{1.0} _{0.6}$ 
&  $6.8 \times 10^{37}$ & 7.5/9 \\
     & Bremsstrahlung & $<$4.2 &  $>$2.6 &\dots 
& $6.1\times 10^{37}$& 7.2/8 \\
     & Bremsstrahlung & 0.056 fixed  &  $>$5.3 &\dots 
& $5.7\times 10^{37}$& 7.2/9 \\
\noalign{\smallskip\hrule\smallskip}
X-10 & Power-law & $<$3.3   &\dots & $1.9 \, \pm \, ^{0.8} _{0.4} $  &  
$1.4 \times 10^{38}$   & 31.2/18\\
    & Power-law      & 0.056 fixed  & \dots & $1.7 \, \pm \, ^{0.2} _{0.3}$ 
&  $1.4 \times 10^{38}$ & 31.9/19 \\
    & Bremsstrahlung & $<$2.0 &   $8.7 \, \pm \, ^{14} _{4.6}$ 
& \dots & $1.4 \times 10^{38}$ &  31.1/18\\
    & Bremsstrahlung & 0.056 fixed  &  $11 \, \pm \, ^{13} _{4.6}$ &\dots 
& $1.3\times 10^{38}$& 31.2/19 \\
\noalign {\smallskip}                       
\hline
\label{tab:pointspectra}
\end{tabular}
\end{table*}

In order to investigate the properties of the other sources identified in
Fig.~\ref{fig:images} MECS data from all 3 observations were used.
Except for M33~X-6 LECS data were not used since there were too 
few counts present. During the 1998 August
observation M33~X-4 was partially obscured by a MECS strongback radial
spoke (see Sect.~\ref{subsect:time}) and these data were excluded from the
analysis.
For all the other sources no obscuration problems were present.
Spectra were extracted centered on each source using the extraction 
radii listed in Table~\ref{tab:pointsources} and 2\arcmin\ in the case
of the LECS M33~X-6 data.
These radii were chosen to minimize 
contamination from nearby sources and take into account the position of
the MECS window support strongback.
The spectra were rebinned in the same way as in Sect.~\ref{sect:obs} and
data selected in the energy ranges where sufficient counts were
obtained.
Since these sources are all faint by BeppoSAX standards care was taken 
to reliably estimate the background. 
In order to properly take into account 
all the possible contaminating contributions, especially from the 
bright nuclear source M33~X-8,  
local backgrounds, each extracted from the same off-axis angle as the 
corresponding source, but at the geometrically opposite position in
the FOV, were used. 
This is also necessary because the MECS instrument background 
has a component due to Mn~K$\alpha$ calibration source X-rays
which are detected within the FOV. This 
effect is most noticeable close to the calibration sources and 
in particular for M33~X-9.
A comparison of these local backgrounds with the average properties
of background taken from high latitude pointings reveals that
no strong evidence for diffuse emission is present in the MECS M33 data.
A higher local background ($\sim$20\%) is present around 
M33~X-8, but this is consistent with being due to the X-ray mirror
scattering wings. 

Instrument response files appropriate to the off-axis positions of each 
source were used, except for M33~X-5 and X-7, which are close 
enough to the on-axis position to use the standard response files.

\begin{figure*}
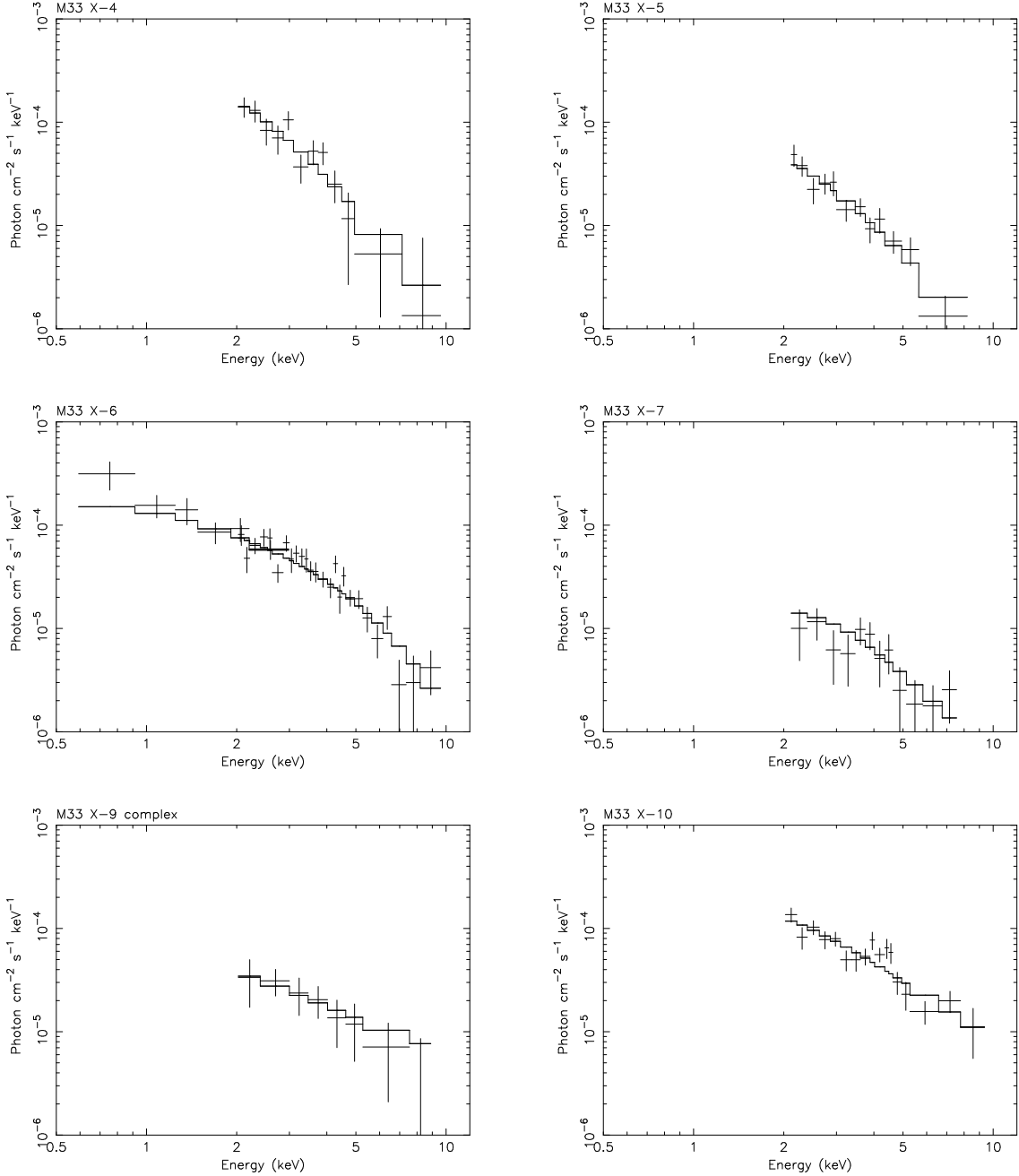


\hbox{\hspace{0.5cm}
\includegraphics[height=6.9cm,angle=-90]{h2503f5a.ps}
\hspace{1.0cm}
\includegraphics[height=6.9cm,angle=-90]{h2503f5b.ps}}
\vbox{\vspace{0.15cm}}
 
\hbox{\hspace{0.5cm}
\includegraphics[height=6.9cm,angle=-90]{h2503f5c.ps}
\hspace{1.0cm}
\includegraphics[height=6.9cm,angle=-90]{h2503f5d.ps}}
\vbox{\vspace{0.15cm}}

\hbox{\hspace{0.5cm}
\includegraphics[height=6.9cm,angle=-90]{h2503f5e.ps}
\hspace{1.0cm}
\includegraphics[height=6.9cm,angle=-90]{h2503f5f.ps}}

\caption[]{MECS photon spectra of the point sources detected
in the M33 galaxy deconvolved using an absorbed power-law model except
M33~X-6 where the LECS spectrum is included and an absorbed disk-blackbody 
model is used. Data from all 3 observations are included, except for 
M33~X-4 where the 1998 August observation is excluded.
The best-fit parameters are given in Table~\ref{tab:pointspectra}. The
plots all have the same extrema to aid comparison}
\label{fig:photons}
\end{figure*}

\subsection{Spectral fit results}

Absorbed power-law and bremsstrahlung models were 
first fit to each of the spectra and the
results presented in Table~\ref{tab:pointspectra}. Fits were
repeated with the absorption constrained to be greater than, and equal
to, the galactic value for each source.
With the exception of M33~X-6 (and possibly M33~X-10)
these models give reasonable fits to all the spectra and it is
not possible to exclude either model.
In the case of M33~X-6, the best-fit absorbed power-law and bremsstrahlung
models give $\chi ^2$'s of 51.7 and 44.2 for 28 dof, respectively.
The best-fit to the M33~X-6 spectrum is instead obtained with an absorbed
disk-blackbody model with ${\rm kT_{in}} = 1.7 \pm 0.2$~keV and 
a normalization (${\rm r_{in} (cosi)^{0.5}}$) of $7 \pm 2$~km.
The 90\% confidence upper limit to a narrow emission line at 6.5~keV
from M33~X-6 is 540~eV. 
In the case of M33~X-10, a broad feature appears to be present at $\sim$4~keV.
This can be modeled by the addition of a ($\sigma = 200$~eV) 
Gaussian line to the best-fit power-law to give a  
$\chi ^2$ of 18.6 for 15 dof. The line energy is
$4.2 \pm ^{0.3} _{0.2}$ keV and the equivalent width 440 eV.

Table~\ref{tab:pointspectra} includes estimates of the source
luminosities extrapolated 
(when necessary) to the 2--10~keV energy range. 
Photon spectra, deconvolved using the best-fit power-law models are shown in 
Fig.~\ref{fig:photons}, except for M33~X-6 where the best-fit 
disk-blackbody model is used. 
With the exception of M33~X-9 and X-10, all the
sources appear to be relatively soft in the 2--10~keV energy range.

\subsection{Time variability}
\label{subsect:time}

\begin{figure*}
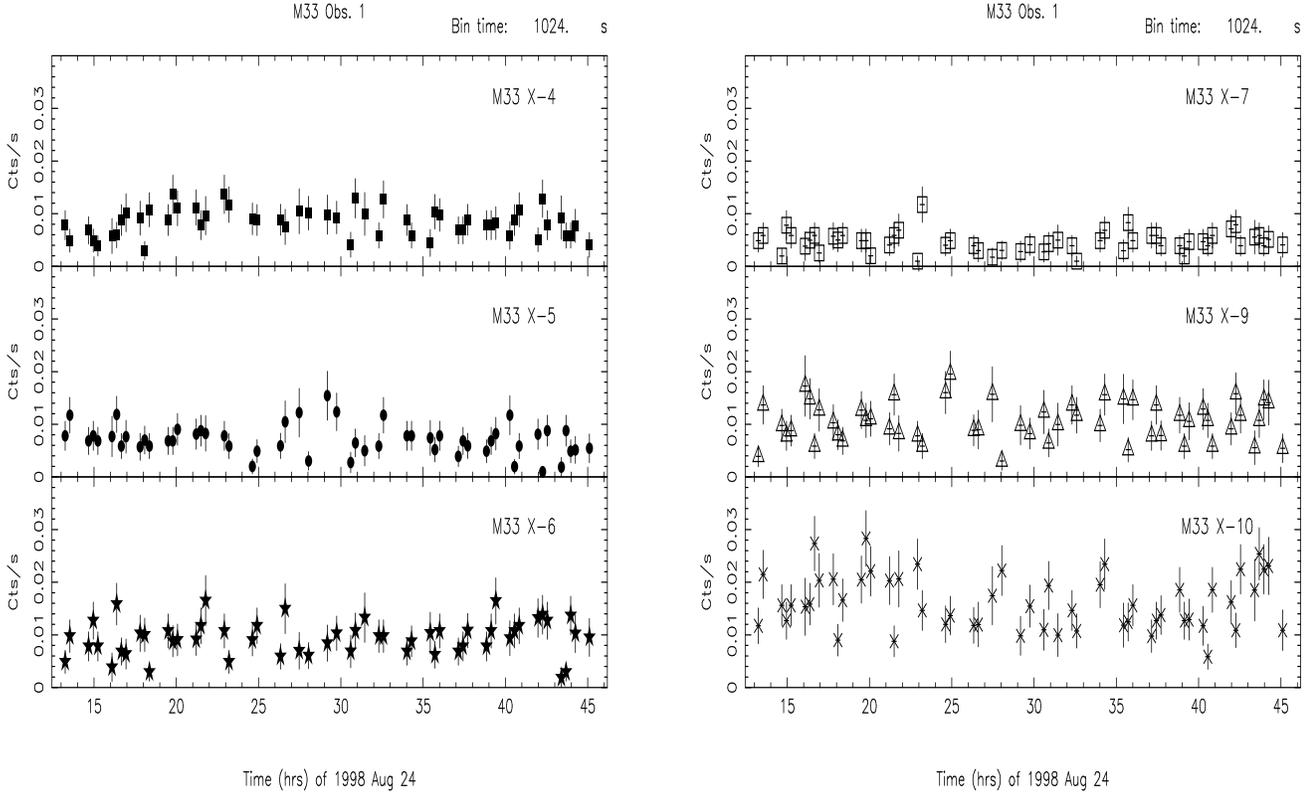

\mbox{ 
\hspace{-0.3cm}
\includegraphics[height=8.0cm,width=10.5cm,angle=-90]{h2503f6a.ps}
\hspace{1.0cm}
\includegraphics[height=8.0cm,width=10.5cm,angle=-90]{h2503f6b.ps}}
\vspace{0.5cm}
\caption[]{Lightcurves for the 6 faint point sources during the 1998 August 
observation. No background has been subtracted}
\label{fig:lc1}
\end{figure*}
 
\begin{figure*}
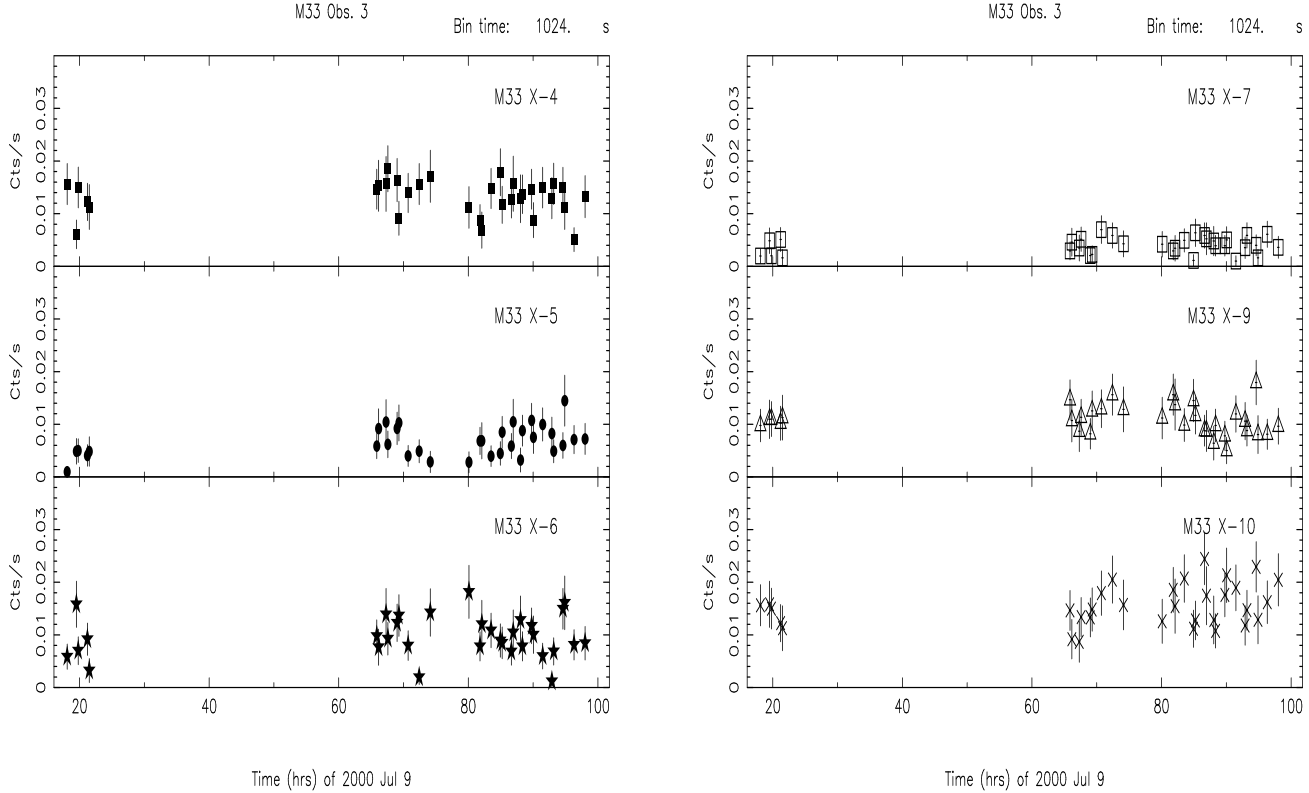

\mbox{ 
\hspace{-0.3cm}
\includegraphics[height=8.0cm,width=10.5cm,angle=-90]{h2503f7a.ps}
\hspace{1.0cm}
\includegraphics[height=8.0cm,width=10.5cm,angle=-90]{h2503f7b.ps}}
\vspace{0.5cm}
\caption[]{Lightcurves for the 6 faint point sources
during the 2000 July observation. No background has been subtracted}
\label{fig:lc3}
\end{figure*}

In order to investigate if any of the point sources exhibit
excess variability, a Kolmogorov-Smirnov test as implemented in 
the {\sc FTOOL}s package {\sc lcstat}
was performed on
each of the lightcurves presented in
Figs.~\ref{fig:lc1} and~\ref{fig:lc3}.
The sources M33 X-4, X-5, and X-10 show some evidence for 
excess variability on time scales $>$1024~s with 
probabilities for being constant of 4.8\%, 4.4\%, and 5.4\%
for M33~X-4, (obs.~1), X-5 (obs.~3), and X-10 (obs.~1), respectively.
The other 3 sources do not show evidence for variability during 
either long observation with probabilities for being
constant of 32\% (X-6, obs.~1), 61\% (X-7, obs.~1, $>$2048~s), 
and 19\% (X-9, obs.~3). 
For comparison, data from source free regions 
of the image were also extracted and processed in the same way.
No evidence for any excess variability was found.

A comparison between the 1998 August 
(Fig.~\ref{fig:lc1}), and 2000 July (Fig.~\ref{fig:lc3}) observations,
does not show evidence for variability, except for
M33~X-4, which showed a factor of $\sim$2 difference in count rate
between the two observations. However, this change 
is almost certainly due to obscuration by one of the radial
spokes of the MECS strongback during the 1998 August observation.
The 2000 July observation was made at a different spacecraft roll
angle, resulting in the source being well clear of the strongback.
The BeppoSAX data are therefore consistent with the 
association of M33~X-4 with the supernova remnant SNR~136+396 discussed 
in Long et al. (\cite{l:96}). 

Table~\ref{tab:var} gives a comparison of the BeppoSAX source intensities
with those derived from previous ROSAT PSPC  
(Long et al. \cite{l:96}) and $Einstein$ IPC observations 
(Peres et al. \cite{p:89}. For M33~X-4 only the data from the 2000
July observation is used.
Unfortunately, due to the lack of good estimates of the absorbing
column to most of the sources (see Table~\ref{tab:pointspectra}),
it is difficult to reliably extrapolate the BeppoSAX spectra
to the PSPC (0.1--2.4 keV) and IPC (0.2--4.5 keV) energy ranges.
Instead, we have adopted the spectral model assumed by 
Long et al. (\cite{l:96}) to convert ROSAT PSPC counts to fluxes.
This consists of a power-law with $\alpha = 2$ and absorption of
$10^{21}$~atom~cm$^{-2}$. We note that none of the spectral
fits reported in Table~\ref{tab:pointspectra} are in
strong disagreement with such a model. We estimate that
the uncertainty in the BeppoSAX fluxes is $\sim$40\%, primarily
due to uncertainties in instrumental calibration at large off-axis
angles.

Examination of Table~\ref{tab:var} shows that M33~X-4, X-6
($Einstein$ only), X-7, X-9 and X-10 (ROSAT only)  
show $>$40\% variability with respect to both the $Einstein$ and ROSAT
measurements. The source M33~X-7 shows the strongest evidence for
variability being a factor $\sim$3 fainter
during the BeppoSAX observations than previously observed. Since
this source exhibits 0.33~day duration eclipses, we verified that the 
observations
were not at the expected times of eclipse using the ephemeris of
Larson \& Schulman (\cite{ls:97}). 
Source X-9, not resolved in the MECS image, appears to be variable being
a factor $\sim$2.5 brighter during the BeppoSAX observation than previously,
if the ROSAT fluxes of the three separate sources are summed together 
(the X-9(c) flux is not reported in Long et al. \cite{l:96}, although it is
visible in their figure and appears to be as bright as the other
two sources). Since the level of variability of M33~X-4, X-6, and X-10 is
comparable with the (hard to quantify) systematic uncertainties, we 
regard the detection of variability for these sources as tentative.  

Since M33~X-4 may be associated with a SNR, it is interesting
to compare \sax\ fluxes with those obtained using $Einstein$ and ROSAT
using a bremsstrahlung model. If  
${\rm N_H}$ is again fixed at $10^{21}$~atom~cm$^{-2}$, we obtain kT=2.4~keV
for the MECS spectrum and 0.1--2.4~keV and 0.2--4.5~keV luminosities of 
1.1$\times10^{38}$~ergs~s$^{-1}$ and 1.5$\times10^{38}$~ergs~s$^{-1}$,
respectively. These values should be compared with 
9.5$\times10^{37}$~ergs~s$^{-1}$ ($Einstein$ IPC, 0.2--4.5 keV) 
and 8.3$\times10^{37}$~ergs~s$^{-1}$ (ROSAT, 0.1--2.4 keV), derived using
this same spectral shape.
Thus, there is no strong evidence for flux variability from M33~X-4.

\begin{table}
\caption{Comparison of the $Einstein$ IPC and ROSAT PSPC,  
luminosities with those derived by extrapolating the BeppoSAX 
values to the $Einstein$ (0.2--4.5 keV) 
and ROSAT (0.1--2.4 keV) energy ranges. See the
text for details of the assumed spectrum.
The absorption corrected luminosity, L, assumes a distance of 795 kpc and
is in units of $10^{37}$ erg~s$^{-1}$}
\label{tab:var}
\begin{tabular}{lcccc}
\hline\noalign{\smallskip}
Source&	\mc{2}{c}{ L (0.1--2.4 keV)} & \mc{2}{c}
{L (0.2--4.5 keV)} \\
      &     BeppoSAX  &   ROSAT &  BeppoSAX & $Einstein$ \\
\noalign{\smallskip\hrule\smallskip} 
X-4    & 18.9 & 11.1&  18.6  & 12.4  \\
X-5    & 5.80 & 5.72&  5.71  & 6.84  \\
X-6    & 14.0 & 18.2&  8.57  & 16.4  \\
X-7    & 2.8  & 12.3&  2.79  & 9.00  \\
X-9    & 8.6  & \dots &  8.57  & \dots  \\
X-9(a) & \dots  & 1.7 &  \dots   & 1.80  \\
X-9(b) & \dots  & 1.9 &  \dots   & 0.67  \\
X-9(c) & \dots  & $-$ &  \dots   & 0.66  \\
X-10   & 29.0 & 18.4 & 23.6  & 16.6  \\ 
\noalign{\smallskip\hrule}
\end{tabular} 
\end{table}

\section{Discussion}
\label{sect:discussion}

The nearby galaxy M33 was observed by the imaging X-ray 
instruments on-board BeppoSAX. As well as the central \src\ 
source which dominates the overall X-ray flux, M33 X-4, X-5, X-6, X-7, 
X-9 and X-10 were clearly detected. 
Observations close to the expected maximum and 
minimum of the 105.9~day intensity cycle of M33~X-8 failed
to reveal the expected modulation, suggesting that it
is probably transitory. No obvious spectral differences between
the observations are present.
The 0.2--10~keV spectrum of \src\ can best be 
modeled by an 
absorbed power-law with $\alpha = 1.89 \, \pm \, ^{0.40} _{0.79}$
and a disk-blackbody with kT $= 1.10 \pm 0.05$~keV and
a projected inner-disk radius of $55.4 \, \pm \, ^{6.0} _{7.7}$~km. 
This spectral shape is in good agreement with the ASCA results
of Takano et al. (\cite{t:94}) and corresponds to a black hole mass
of $\sim$10~M$_\odot$. 

The nature of \src\ is intriguing. X-ray transients
such as Aql~X-1 and 4U~1630-47 show outbursts on
timescales of hundreds of days, but are not periodic (e.g., 
Kuulkers et al. \cite{k:97}). A number of persistent
X-ray sources exhibit long term periodicities in their 
X-ray intensities. Perhaps the best known of these
are the pulsars Her~X-1 (35 days, Giacconi et al. \cite{g:73}),
LMC~X-4 (30.5 days, Lang et al. \cite{l:81a}) and SMC~X-1
(60~days, Wojdowski et al. \cite{w:98}). In addition, the
high-mass X-ray binary SS~433 thought to contain a black hole
exhibits a 164~day photometric and spectroscopic modulation in
the optical band (Margon et al. \cite{m:79}). Two other
black hole candidates, Cyg~X-1 (300 days, Priedhorsky et al. 
\cite{p:83} and LMC~X-3, (198 or 99 days, Cowley et al. \cite{c:91})
also exhibit long term X-ray modulations. Thus, the presence
of a 105.9~day X-ray intensity modulation from \src\ is not
particularly unusual if the source does indeed contain 
a $\sim$10~M$_\odot$ black hole. 

Recently, extended observations using the ASM on R-XTE have 
revealed that in at least four sources (Cyg~X-3, LMC X-3, 
SMC X-1, and GX~354-0; Paul et al. \cite{p:00}; Wojdowski et al.
\cite{w:98}; Kong et al. \cite{k:98}) the long term
modulations have disappeared, or changed periods. Such unstable
behavior is difficult to reconcile with the preferred model
for the modulations in the high-mass sources which is obscuration by
a precessing accretion disk (see e.g., Priedhorsky \& Holt \cite{ph:87}).
Since the detection of changes in the properties of these
long term variations requires very long observing
baselines, it is possible that such unstable behavior is relatively common
in accreting systems. Thus, the probable cessation
of the 105.9~day periodicity reported here strengthens somewhat the
idea that \src\ consists of a black hole accreting from a 
binary companion.

The 2--10~keV spectra of M33~X-4, X-5, X-7, X-9 and X-10 
are all consistent with power-law or bremsstrahlung models, whilst that 
of X-6 appears to be significantly more complex, and may be reasonably well 
modeled with a disk-blackbody with kT = $1.7 \pm 0.2$~keV and
a projected inner-disk radius of $7 \pm 2$~km. 
This normalization is in good agreement with the average of a number
of low-mass X-ray binary systems known to contain neutron stars ($\sim$7~km),
but is smaller than the average ($\sim$25~km)
for black hole candidates (e.g., Tanaka \& Lewin (\cite{tl:95}). 
This suggests that M33~X-6 may be a neutron star accreting from a 
low-mass companion.
The spectrum
of M33~X-9 is quite hard with $\alpha = 1.3$, similar to 
the hard/low state in galactic black hole candidates. 
A comparison with earlier $Einstein$ and ROSAT results implies that 
M~33 X-7 and X-9 are variable on long timescales, M33~X-4,
X-6 and X-10 may be variable, 
whilst X-5 does not show evidence for any such variability.
We cannot exclude the association of M33~X-4 
with the supernova remnant SNR~136+396 discussed 
in Long et al. (\cite{l:96}). 
The lack of variability, moderate luminosity 
($3 \times 10^{37}$~erg~s$^{-1}$), and simple (consistent with a 
power-law or bremsstrahlung) spectrum may indicate that 
M33~X-5 is a young X-ray bright supernova remnant.
M33~X-6 clearly has an unusual spectrum and is worthy of further study
with more sensitive instruments.
It is unsurprising that M33~X-7 shows strong long term variability since
it is known to be an (eclipsing) X-ray binary system
(Peres et al. \cite{p:89}). Whilst M33~X-9 was a factor $\sim$2.5
brighter during the BeppoSAX observations than previously observed,
it is unclear which (if any) of the 3 sources that apparently comprise
X-9 had become significantly brighter. 

\begin{acknowledgements}
The \sax\ satellite is a joint Italian-Dutch programme.
L. Sidoli acknowledges an ESA Research Fellowship. 
We thank the staff of the \sax\ Science Data Center
for help with scheduling these observations and M. Corcoran for 
assistance with the ROSAT analysis. The referee, H.L. Marshall, is
thanked for helpful comments.
\end{acknowledgements}


\begin{thebibliography}{}

\bibitem[1997]{b:97}
Boella G., Chiappetti L., Conti G., et al., 1997, A\&AS 122, 327

\bibitem[1974]{b:74}
Boulesteix J., Courtes G., Laval A., et al., 1974, A\&A 37, 33 

\bibitem[1982a]{cs:82a}
Christian C.A., Schommer R.A., 1982a, ApJ 253, L13

\bibitem[1982b]{cs:82b}
Christian C.A., Schommer R.A., 1982b, ApJS 49, 405

\bibitem[1991]{c:91}
Cowley A.P., Schmidtke P.C., Ebisawa K., et al., 1991, ApJ 381, 526 

\bibitem[1990]{d:90} 
Dickey J.M., Lockman F.J., 1990, ARA\&A 28, 215

\bibitem[1997]{d:97}
Dubus G., Charles P.A., Long K.S., Hakala P.J., 1997, ApJ 490, L47

\bibitem[1999]{d:99}
Dubus G., Charles P.A., Long K.S., Hakala P.J., Kuulkers E., 1999, MNRAS, 
302, 731

\bibitem[1997]{f:97}
Frontera F., Costa E., Dal Fiume D., et al., 1997, A\&AS 122, 371

\bibitem[1973]{g:73}
Giacconi R., Gursky H., Kellogg E., et al., 1973, ApJ 184, 227

\bibitem[1987]{g:87}
Gottwald M., Pietsch W., Hasinger G. 1987, A\&A 175, 45

\bibitem[2000]{hp:00}
Haberl F., Pietsch W., 2000, in The Interstellar Medium in M31 and
M33, eds. E.M. Berkhuijsen R. Beck and R.A.M. Walterbos
(Aachen: Shaker Verlag)

\bibitem[1998]{k:98}
Kong A.K.H., Charles P.A., Kuulkers E., 1998, New Astronomy 3, 301

\bibitem[1997]{k:97}
Kuulkers E., Parmar A.N., Kitamoto S., Cominsky L.R., 
Sood R.K., 1997, MNRAS 291, 81 

\bibitem[1981]{l:81a}
Lang F.L., Levine A.M., Bautz M., et al., 1981, ApJ 246, L21

\bibitem[1997]{ls:97}
Larson D.T., Schulman E., 1997, AJ 113, 618 

\bibitem[1981]{l:81}
Long K.S., D'Odorico S., Charles P.A., Dopita M.A., 1981, ApJ 246, L61

\bibitem[1996]{l:96}
Long K.S., Charles P.A., Blair W.P., Gordon S.M., 1996, ApJ 466, 750

\bibitem[1986]{m:86}
Makishima K., Maejima Y., Mitsuda K., et al., 1986, ApJ 285, 712           

\bibitem[1983]{m:83}
Markert T.H., Rallis A.D., 1983, ApJ 275, 571

\bibitem[1984]{m:84}
Mitsuda K., Inoue H., Koyama K., et al., 1984, PASJ 36, 741            

\bibitem[1979]{m:79}
Margon B., Grandi S.A., Stone R.P.S., Ford H.C., 1979, ApJ 233, L63

\bibitem[1983]{mc:83}
Morrison D., McCammon D., 1983, ApJ 270, 119

\bibitem[1997]{p:97} 
Parmar A.N., Martin D.D.E., Bavdaz M., et al., 1997, A\&AS 122, 309

\bibitem[2000]{p:00}
Paul B., Kitamoto S., Makino F., 2000, ApJ 528, 410

\bibitem[1989]{p:89}
Peres G., Reale F., Collura A., Fabbiano G., 1989, ApJ 336, 140

\bibitem[1987]{ph:87}
Priedhorsky W.C., Holt S.S., 1987, Space Sci. Rev. 45, 291

\bibitem[1983]{p:83}
Priedhorsky W.C., Terrell J., Holt S.S., 1983, ApJ 270, 233

\bibitem[1993]{s:93}
Schulman E., Bregman J.N., Collura A., Reale F., Peres G., 1993, ApJ 418, L67 

\bibitem[1994]{s:94}
Schulman E., Bregman J.N., Collura A., Reale F., Peres G., 1994, ApJ 426, L55 

\bibitem[1979]{s:79}
Sunyaev R.A., Tr\"umper J., 1979, Nat 279, 506

\bibitem[1994]{t:94}
Takano M., Mitsuda K., Fukazawa Y., Nagase F., 1994, ApJ 436, L47

\bibitem[1995]{tl:95} 
Tanaka K., Lewin W.H.G., 1995, in X-ray Binaries, eds. W.H.G.
Lewin J. van Paradijs and E.P.J. van den Heuvel (Cambridge: Cambridge
Univ. Press), p.~121

\bibitem[1988]{t:88}
Trinchieri G., Fabbiano G., Peres G., 1988, ApJ 325, 531

\bibitem[1991]{vdb:91}
Van den Bergh S., 1991, PASP 105, 609

\bibitem[1998]{w:98}
Wojdowski P., Clark G.W., Levine A., et al., 1998, ApJ 502, 253

\end{thebibliography}
\end{document}